\documentclass[11pt]{article}

\usepackage[cmtip,arrow]{xy}
\usepackage{pb-diagram,pb-xy}

\usepackage{makeidx}         % allows index generation
\usepackage{graphicx}        % standard LaTeX graphics tool
\usepackage{multicol}        % used for the two-column index
\makeindex             % used for the subject index
\newlength{\vshift}
\newlength{\hshift}
\usepackage{amssymb,amsopn}

\begin{document}
\begin{titlepage}
\rightline{}
\rightline{}
\vspace{2em}
\begin{center}
{\huge {\bf Generalized Path Dependent Representations for Gauge Theories}}
\vskip 3em
{{\sc \bf Marat C. Reyes}}
\vskip 2em
{ Instituto de Ciencias Nucleares, Universidad Nacional
Aut\'onoma de M\'exico, 04510 M\'exico D.F., M\'exico  
and Universidad 
T\'ecnica Federico Santa Maria, Campus Santiago, C.P. 766-0251, Santiago, Chile.}
\end{center}
\begin{abstract}
A set of differential operators acting by continuous
 deformations on path dependent functionals of open and closed curves
is introduced. Geometrically, these path operators are 
interpreted as infinitesimal generators of curves 
in the base manifold of the gauge theory. They furnish a 
representation with the action of the group of loops having
 a fundamental role. 
We show that the path derivative, which is covariant by 
construction,
satisfies the Ricci and Bianchi identities. Also, we provide a geometrical derivation
of covariant Taylor expansions based on particular deformations of open curves. The formalism 
includes, as special cases, other path dependent operators such as 
end point derivatives and area derivatives.
\end{abstract}
\vspace*{0.5cm}
%{\bf Keywords:} d, 
%PACS: 02.40.Gh, 02.20.Uw\\
%\hspace*{0.6cm}MSC: 81T75 
\vspace*{0.5cm}
\vfill
\quad\scriptsize{email: carlos.reyes@nucleares.unam.mx } 
\vfill
\end{titlepage}\vskip.2cm
\newpage
\setcounter{page}{1}
\newcommand{\Section}[1]{\setcounter{equation}{0}\section{#1}}
\renewcommand{\theequation}{\arabic{section}.\arabic{equation}}
\Section{Introduction}%.........................................................................Introduction
The importance attributed to the notion of path dependence 
goes back to Dirac's work on the 
non-integrability of the phase exhibited by 
wave functionals in the presence of electromagnetic fields 
in Quantum Mechanics \cite {Dirac}. Since then, the idea 
of incorporating path dependence in gauge formulations, including 
gravity, has been explored and studied in 
various physical contexts \cite{Schwinger,Mandelstam,Yang,
Polyakov,Gambini-loop formulation}. 
In the framework of gauge theory it
was originally
Mandelstam who elaborate 
a gauge invariant 
formulation of electrodynamics coupled to a scalar field
 with a definite open path dependence \cite{Mandelstam}.
Extensions for the non-Abelian case were developed in \cite {Bia}. 
The global and geometrical aspects of non abelian gauge 
theories were highlighted in
the integral formulation of Yang \cite {Yang} and Wu-Yang \cite
 {Wu and Yang}. 
Moreover, the loop representation \cite{Gambini-loop 
formulation, Smolin} based on closed curves is one
essential tool in the loop quantum gravity approach.

However, and in spite of the advantages that may present 
path dependence in gauge theories 
we believe there is still a gap to appropriately understand 
in which manner the existent path dependent
operators are related. The tendency until now has been to 
adapt each definition of path derivative to a specific 
domain of considerations which are believed to be relevant 
for the construction. In general all
these considerations result to be different \cite{Loll}. 
Efforts to study end point derivatives and area derivatives 
from a rigorous mathematical 
viewpoint had been carried out in \cite{Tavares}, however a 
different definition for both path derivatives is formulated.
Several definitions of path dependent operators can be found 
in \cite{Nambu,Polyakov,Makeenko,Chan,Corrigan-Durand}. 
The definition of path dependent operators had been made 
essentially depending on: 

i) The space where path dependent functionals take values 
is either the space of open or closed curves and if the space
includes base points or not.

ii) The nature of the variation is due to a point or many 
points, which have been 
usually called end point derivatives and area derivatives 
respectively.

iii) The place where the variation is appended, is on the 
curve or in other place on the manifold. 

In this paper, we address the three points above. 
We introduce a covariant path derivative acting by 
continuous deformations
on a general class of open and closed curves. We show
that the definition of path derivative 
generalizes all types of
end points and area derivatives. Geometrically, the path 
derivative is an infinitesimal generator 
of curves which under some assumptions enables a 
representation with the action of the 
group of loops deeply involved. We show that the path 
derivative 
satisfies the Ricci and Bianchi identities. The work is 
organized as follows. In section 2
we define the path derivative emphazising the role of the 
group of loops. Next, in the third section we compare the direct consequences
of the action of the path operator with well known equations, 
this  amounts to compute the path derivative of phase factors and 
scalar fields. Area and end point derivatives are identified and 
related therein. In section 4 
we calculate the finite variation of a functional when its argument
is changed by successive infinitesimal deformations. This change 
may be interpreted through the action of the group of loops on 
arbitrary paths which we represent by a set of differential 
operators. 

In section 5 we use the path derivative to obtain 
covariant Taylor series. These series arise when using particular deformations 
which do not enclose area 
and collapse to a point in the original curve, so only
open curves can be considered.

Finally we discuss some aspects related to the loop derivative 
defined in \cite{Gambini-loop formulation}. The loop derivative
emerges in our approach when we restrict to spatial curves and 
to deformations with end points fixed.
\section{The Path Derivative}%................................................................The path derivative
The gauge theory is introduced by considering the principal 
fiber $P(G, M)$ with
a one form connection $A$ valued in the algebra of the gauge
 group $G$. We will denote by $\Gamma(M)$ the space of all
open and closed smooth curves in the base manifiold $M$ 
equipped with the usual properties of path composition 
\cite{Gambini-Pullin}.
Let us consider a general class of path-dependent matrix
 functionals $\Psi$ taking values in the space $\Gamma(M)$ 
and transforming covariantly under the gauge group $G$. 

We define the path derivative of the functional $\Psi(\alpha)$ 
for a given path $\alpha \in \Gamma(M)$ by

\begin{eqnarray}\label{derivative}
\mathcal D \Psi(\,\alpha)=\Delta \Psi(\alpha)\,
 -\Psi(\alpha),
\end{eqnarray}
\noindent such that the action $\Delta:\Psi(\alpha) \to 
\Psi'(\alpha')$, is to displace infinitesimally and continuously 
the initial curve $\alpha$
to a deformed curve $\alpha'$ with some transforming action
 on $\Psi$ specified below. 
To begin the construction we consider the same parametrization for 
the curves $\alpha(\sigma)$ and $\alpha'(\sigma)$
 with $\sigma \in [0,1]$, and corresponding end 
points $(x,y$) and $(x',y')$ as indicated in Fig 1.
 Also, let us represent the trajectories 
followed by the points of the initial curve $\alpha(\sigma)$
 along the deformation
as the family of diffeomorphism defined by
$x^{\mu}(\sigma,t)=x^{\mu}(\sigma)+ \delta x^{\mu}(\sigma,t)$
 and parameterized with $t \in [0,1]$. The 
initial and final curves being respectively,
$x^{\mu}(\sigma,0)=x^{\mu}(\sigma)$ and $x^{\mu}
(\sigma,1)=x'^{\mu}(\sigma)$. 

We adopt the view that the operator $\mathcal D$
generates a
vector field $\vec N (\sigma, t)$ with
\begin{eqnarray}\label {int curves}
 N^{\mu}(\sigma,t)= \frac{ \partial x^{\mu}(\sigma,t)}{ \partial t},
\end{eqnarray}
\noindent and where $x^{\mu}(\sigma,t)$ are the integral
 curves associated to the deformation of $\alpha(\sigma)$. 
 With this
in mind 
$\mathcal D$ will be denoted alternatively by $\mathcal D(N)$.
\begin{figure}
\begin{center}
\includegraphics[width=6cm]{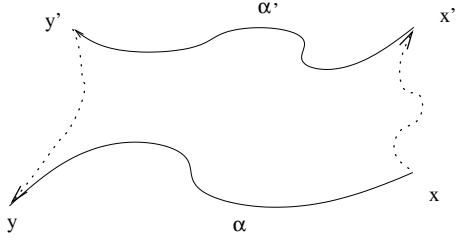}
\caption{Deformation of the curve $\alpha \to \alpha'$.}
\end{center}
\end{figure}
We assume a transformation of a matrix functional $\Psi_{AB}
(\alpha)$ 
under the action of the deformation by
\begin{eqnarray}\label{representation}
\Delta_t \Psi_{AB}(\alpha)=U_{A}^{A'}(\delta y^{-1}) \,  
\Psi_{A'B'}(\,  \alpha' 
)  \, U_{B}^{B'}(\delta x),
\end{eqnarray}
\noindent where the elements $U_ {B}^{B'}(\delta x)$ and
 $U_{A}^{A'}(\delta y^{-1})$ transform internal indices
$ \{A,B,\dots \}$ to one system to another and are functions 
of the paths 
which connect the end points of both curves $\alpha$ and $\alpha'$, 
see Fig 1. The notation introduced is 
$\delta x(0,t)=\delta x$ and $\delta y^{-1}$ for $\delta x(1,t)=\delta y$ 
traversed in the
opposite direction. To ensure covariance in the definition
 (\ref{derivative}) we define the $U$
objects to transform covariantly and for the property 
$\Delta_t\, \Delta_{t'}=\Delta_{t\,o\,t'}$ to hold,
their composition rules
\begin{eqnarray}\label{composition}
U_{A'}^{A''}(\delta x')\, U_{A}^{A'}(\delta x)=U_{A}^{A''}
(\delta x \,o \,\delta x'),\\
U_{B'}^{B''}(\delta y')\,U_{B}^{B'}(\delta y)=U_{B}^{B''}
(\delta y \,o \,\delta y'),\nonumber
\end{eqnarray}
\noindent which is the reflection in functional space of
 the geometric property of path composition.
This suggest to identify $U$ with ordered phase factors 
of
the local connection $A$. 

Phase factors are defined for a given path $\gamma$,
\begin{eqnarray}
U(\gamma)= \mathcal{P}_s\left(\, \exp \int -A_{\mu}(s)\, 
 \frac{d \gamma^{\mu}(s)}{ds} ds \, \right),
\end{eqnarray}
\noindent such that $\mathcal P_s$ means path ordered. 
The usual composition properties are,
\begin{eqnarray}
U(\gamma')\,U(\gamma'')=U(\gamma' \,o \,\gamma''),
\\ U(\gamma)\, U(\gamma^{-1})=1, \label{retraced}
\end{eqnarray}
\noindent where the path composition $\gamma=\gamma' \, 
o\,\gamma''$ is to be read with the convention that takes
 $\gamma'$ followed by $\gamma''$.
Now, the phase factors in (\ref{representation}) 
approximated to the first order in $N (\sigma,t)$ are
\begin{eqnarray}\label{prop1}
 U(\delta x)=1
- \int_0^1 dt \,N^{\mu}(0,t)\,A_{\mu}(x(t)), \nonumber \\  U(\delta y^{-1})=1
+ \int_0^1 dt \, N^{\mu}(1,t)\,A_{\mu}(y(t)).
\end{eqnarray}
From (\ref {prop1}), (\ref{representation}) and (\ref {derivative}) 
we write for any matrix path dependent quantity $ \Psi_{y, x}$
connecting the points $x$ and $y$, not necessarily distinct,
\begin{eqnarray}\label{variation}
\delta \Psi_{y,x}&=&\Psi_{y,x}\int_0^1 \, N^{\mu}(0,t) A_ {\mu}(x(t))
 - \int_0^1 N^{\mu}(1,t) A_ {\mu}(y(t))\, \Psi_{y,x}\nonumber \\&&
+\, \int _0^1 \int_0^1 d\sigma dt \,   N^{\mu}(\sigma,t)\, \mathcal 
D_{\mu} (\sigma,t)\Psi_{y,x},
\end{eqnarray}
\noindent where  
$\delta \Psi= \Psi(\, \alpha'
)-\Psi(\alpha)$ and the definition of the functional derivative 
$\mathcal D_ {\mu}(\sigma,t)$ is through the expression
\begin{eqnarray}\label{func-der}
\mathcal D(N)=\int _0^1 \int_0^1 d\sigma dt \,   
N^{\mu}(\sigma,t)\, \mathcal D_{\mu} (\sigma,t).
\end{eqnarray}
Note that for the particular choice $x^{\mu}(\sigma,t)=x^{\mu}
(\sigma)+ t\,\delta x^{\mu}(\sigma)$ which considers straight paths
connecting both curves, the equation (\ref{variation}) reduces 
to one obtained in reference \cite{Corrigan-Durand}. 
It is the generality 
of equation (\ref{derivative}) together with the geometrical 
and global character of (\ref{representation}) that allow for the
generalization introduced here and the use
of the group of loops in the underlying structure of equations 
exhibited further. 

Now, for the curve deformation that just moves one point along a straight line,
 which we call a point deformation, we define, for example for the end 
point $y$,
\begin{eqnarray} \label{var y}
\mathcal D_{\delta y} \Psi_{y,x}=\delta \Psi_{y,x}+ \delta  y^{\mu} A_ {\mu}(y)  \, \Psi_{y,x}, \,
\end{eqnarray}
\noindent analogously for the initial point $x$
\begin{eqnarray}
\mathcal D_{\delta x} \Psi_{y,x}=\delta \Psi_{y,x}-\Psi_{y,x} \, \delta  x^{\mu} A_ {\mu}(x).
\end{eqnarray}
And for the curve deformation with $x$ and $y$ fixed but that encloses some area 
we define the loop deformation by
\begin{eqnarray}\label{loop deformation}
\mathcal D_{L} \Psi_{y,x}=\delta \Psi_{y,x}.
\end{eqnarray}
As mentioned before the construction can be understood in terms
of the action of the group of loops $L$ on arbitrary paths $\gamma$
belonging to $\Gamma(M)$. The action defined by usual path composition. Let us consider the same path $\alpha(\sigma)$ as before and
focus on the loop $l=\delta x \, o \, \alpha' \, o \, \delta y^{-1 }\,o \, \alpha^{-1}$ 
with composition $ l \, o\, \alpha=\delta x \, o \, \alpha' \, o \, \delta y^{-1 }$. 
Also, the variation of a functional
 $\Delta \Psi(\alpha)$ will be represented by an operator $U(l)$ with $l\in L$ as,
\begin{eqnarray}\label{rep-loop}
\Psi( l \, o\, \alpha )= U(l) \Psi(\,  \alpha
) ,
\end{eqnarray}
\noindent and therefore given (\ref{representation}) we have
\begin{eqnarray}\label{loop}
\Psi(\alpha')=U(\delta y) \, \Big[ \,U(l)\, \Psi(\,  \alpha
) \,\Big] \, U(\,\delta x^{-1})\,.
\end{eqnarray}
In the next sections we justify eq (\ref{rep-loop}) and we give 
a precise meaning to the operator $U(l)$ in terms of the operator $\mathcal D$.

Moreover, we see that the expression
\begin{eqnarray}\label{comp}
\Psi( l \, o\, \alpha )=U(\delta y^{-1})\,  \Psi(\alpha')\,U(\delta x),
\end{eqnarray}
\noindent behaves as expected with respect to the composition 
of loops. To see this, consider the action of the two loops $l_1=  \,\delta x_1 \,
o \,\alpha' \,o  \,\delta y
^{-1}_1 \,o\,\alpha^{-1}$ and $l_2= \delta x_1 \,o\,\delta x_2 \,
o \,\alpha''\,o\,\delta y
^{-1}_2 \,o\,\delta y_1^{-1}\,o\, \alpha'^{-1}$ on $\alpha$ which gives
$ l_2 \,o \,l_1\,o\,\alpha=\delta x_1 \,o \,\delta x_2 \,
o \,\alpha'' \,o \,\delta y^{-1}_1 \,o \,\delta y
^{-1}_2$. Thus eq (\ref{comp}) is with respect to the composition of loops 
\begin{eqnarray}
\Psi( l_2 \,o \, l_1 \, o\, \alpha )=U(\delta y^{-1}_1 \,o \,\delta y^{-1}_2   )\, 
 \Psi(\alpha'')\,U(\delta x_1 \,o\, \delta x _2 ),
\end{eqnarray}
\noindent where (\ref{composition}) and (\ref{retraced}) have been used.
\section{Covariant Differentiation of Gauge Objects}%...........................................................Differentiation
Here we compute the action of the path derivative on 
variables arising in gauge theories such as phase factors, 
path dependent matter fields and local gauge fields. 
At the end of the section we provide a relation between 
the path derivative introduced here and the so called end point
 and area derivatives. 

Let us consider the ordered phase factor of the same path as 
before $\alpha(\sigma)$,
\begin{eqnarray}
U_ {y, x}(\alpha)= \mathcal{P}_{\sigma}\left(\, \exp \int_0 ^1 
 -A_{\mu}(\sigma) \frac{d \alpha^{\mu}(\sigma)}{d\sigma}\,d\sigma \, \right),
\end{eqnarray} 
\noindent from definition (\ref {derivative}) and (\ref{representation}), 
\begin{eqnarray}\label{derphase}
\mathcal D U(\,\alpha)= U(\delta y^{-1}) \,  U(\,\alpha') \, 
 U(\delta x)-U(\,\alpha). 
\end{eqnarray}
We partition the paths $\alpha$ and $\alpha'$ in $N$ segments. 
Each segment $\alpha_{i+1,i}$ defines a 
phase factor joining the points $x_i=x(\sigma_i)$  
and $x_{i+1}=x(\sigma_{i+1})$, similarly for $\alpha'$. 
By the composition property 
of phase factors we have  
\begin{eqnarray}\label 2
U(\alpha')=\prod_{i=0}^{N}U(\alpha'_{i+1,i})=U'_{N+1,N}\, \dots  \,U'_{2,1}\, U'_{1,0},
\end{eqnarray}
\noindent and considering each segment $\alpha'_{i+1,i}=
\delta x_i^{-1} \,o\, l_i \,o\, \alpha_i \,o\,  \delta x_{i+1}$
in terms of the loop $l_i=\delta x_i \, o \, \alpha' \, 
o \, \delta x_{i+1}^{-1 }\,o \, \alpha^{-1}$ we may write
\begin{eqnarray}\label {1}
U(\alpha'_{i+1,i})=U(\delta x_{i+1})\, U(\alpha_{i+1,i})\,U(l_i) 
\,U(\delta x_i^{-1}).
\end{eqnarray}
Therefore from eqs (\ref{1}) and (\ref{2}), $\mathcal DU(\,\alpha)$ is written
as a product of paths given by 
\begin{eqnarray}
\mathcal DU(\,\alpha)= \prod_{i=0}^{N}  \Big( U(\alpha_{i+1,i})\, H(x_i)\Big) \, 
-U(\alpha),
\end{eqnarray}
\noindent where we have replaced $U(\,l_i)$ by the holonomy 
$H(x_i)$ evaluated along the line $x_i=x(\sigma_i,t)$. Now,
$H(x_i)$ can be written using the non abelian Stokes theorem to lowest order
as
\begin{eqnarray}
H(x_i)=1- \int_0^1 \mathcal F_{\mu \nu}(x_i) \, N^{\mu}(\sigma_i,t)
 \frac{\partial x^{\nu}_i }{\partial \sigma} d\sigma _i dt,
\end{eqnarray}
\noindent where $\mathcal F_{\mu \nu}(x_i)=U(\delta x_{i})
F_{\mu \nu}(x_i)U(\delta x^ {-1}_{i})$ is the parallel 
transported curvature,
see \cite{etal,Stokes}. Replacing, we have
\begin{eqnarray}
\mathcal D(N)\, U(\,\alpha)&=& - \int_0^1\sum_{i=0}^{N}
 U_{i+1,i}\, \mathcal F_{\mu \nu}(x_i)
\,  N^{\mu}(\sigma_i,t)\, \frac{\partial x_i^{\nu} }{\partial \sigma} \,d\sigma _i\, dt.
\end{eqnarray}
The continuum limit of the above equation gives
\begin{eqnarray}
\mathcal D(N) \,U(\alpha)=-\int_0^1dt \int_{0}^{1} d\sigma 
\, U_{y,x(\sigma,t)}\, \mathcal F_{\mu \nu}(x(\sigma,t))\, U_{\,x(\sigma,t),x}\, N^{\mu}(\sigma,t) \,  
\frac{\partial x^{\nu}(\sigma,t) }{\partial \sigma}.
\end{eqnarray}
It is easy to show that the result of applying the path 
derivative on phase factors gives the usual covariant derivative on the path
introduced in \cite{Corrigan-Durand} if
 both curves are connected by straight line segments. 
 From the definition (\ref{func-der}) we read
\begin{eqnarray}
\mathcal D_{\mu}(\sigma,t)\, U(\alpha)= -U_{y,x(\sigma,t)}
 \,\mathcal F_{\mu \nu}(x(\sigma,t))\,U_{x(\sigma,t),x}  \frac{\partial x^{\nu}(\sigma,t) }{
\partial \sigma}.
\end{eqnarray}
It is straightforward to compute the point deformation
 of the phase factor $U(\gamma)$
that changes $z$ to $z+\delta z$. This is
\begin{eqnarray}
\mathcal D_{\delta z} U_{AB}(\gamma)=0,
\end{eqnarray}
\noindent which is consequence of using the retracing 
property (\ref{retraced}). Analogously, 
the point deformation on $x$ of the path dependent 
 scalar field
defined by
$\Phi_{AB}(\gamma,x)= U_{AC}(\gamma)\,\phi_{CB}(x)$ is
\begin{eqnarray}\label{Mandelstam}
\mathcal D_{\delta x} \Phi_{AB}(\gamma,x)&=&U_{AC}(\gamma)
\Big(U_{CD'}(\delta x^{-1}) \phi_{D'E'}(x')U_{E'B}(\delta x)-
 \phi_{CB}(x) \Big) \nonumber \\
&=&U_{AC}(\gamma)\delta
 x^{\mu}  \nabla_{\mu}\phi_{CB}(x),
\end{eqnarray}
\noindent where
 $\nabla_{\mu}=\partial_{\mu}+[\,A_{\mu},~~]$ is the usual 
 covariant derivative defined on point functions. We emphasize 
 that although the action of $\mathcal D_{\delta x}$ is the 
 same of the end point derivative on path dependent 
 scalar fields, their definitions are formulated rather different.

In the same way the action of the derivative on the local gauge field
$A_{\mu}(x)$ is
\begin{eqnarray}
(\mathcal D_{\delta x} A_{\mu})_{AB}&=&\left( \delta x^{\alpha}
 \partial_{\alpha}A_{\mu}+\delta x^{\alpha} [A_{\mu},A_{\alpha}]  \right)_{AB},
\end{eqnarray}
\noindent where we have considered an arbitrary path 
contracted to the point $x$. From here we have
\begin{eqnarray}
\delta x^{\mu}\mathcal D_{\delta x} A_{\mu}(x)&=& \delta x^{\mu}
 \delta x^{\alpha} \partial_{\alpha}A_{\mu} (x),  
\end{eqnarray}
\noindent which is just the directional derivative of the gauge field.
\section{The Path Derivative as the Generator of Curves}%....................................Generator
Eq
(\ref{rep-loop}) defines a representation based on 
differential operators associated to a family of deformed curves.
Let us proceed to explicitly calculate these operators 
and relate them to the action of the group of loops. 
 
Consider the functional $\Psi(\alpha(\sigma))
$ and a
finite variation of the path $\alpha(\sigma) \to \alpha'
(\sigma)$. Using the same notation as before we introduce a family 
of deformed curves $\alpha_t(\sigma)$ parametrized with $t\in [\,0,1\,]$. For an 
$N$ partition of the $t$-interval $[\,0,1\,]$ the
infinitesimal deformations are 
\begin{eqnarray}  \label{3}
 \Psi(\alpha_{n+1})= \Psi(\alpha_{n})+\Psi(\alpha_{n}) 
  A_{x_{n}}  -A_{y_{n}} \Psi(\alpha_{n})+ {\mathcal D} (N_{n}) \Psi(\alpha_{n}),  
\end{eqnarray}
\noindent with the definitions,
\begin{eqnarray}  
A_{x_{n}} =\int_{t_{n}}^{t_{n+1}}dt\, N^{\mu}(0,t) \,A_{\mu}(x(t)), \\ 
A_{y_{n}} =\int_{t_{n}}^{t_{n+1}}dt\, N^{\mu}(1,t)\, A_{\mu}(y(t)), \\ 
\mathcal D(N_{n})=\int_{t_{n}}^{t_{n+1}}dt \int_0^1 d\sigma\,  
 N^{\mu}(\sigma,t)\mathcal D_{\mu}(\sigma,t),
\end{eqnarray}
\noindent and $\alpha_n=\alpha_{t_n}(\sigma)$ together with 
$x=\alpha_1 (0)$, $x'=\alpha_N (0)$, and $y=\alpha_1 (1)$, $y'=\alpha_N(0)$. 
Iterating  $n$ times the equation (\ref{3}) and using the identities 
\begin{eqnarray}
U_{x',x}=\lim_{N\to \infty}(1-A_{x_N})\,(1-A_{x_{N-1}})\dots (1-A_{x_1}),
\end{eqnarray}
\begin{eqnarray}
U^{-1}_{x',x}=\lim_{N\to \infty}(1+A_{x_1})\dots (1+A_{x_{N-1}})\, (1+A_{x_N}),
\end{eqnarray}
\noindent and 
\begin{eqnarray}
\int_0^1 dt\int_0^t dt'\,f(t,t') +\int_0^1 dt'\int_0^{t'}
 dt \,f(t,t')=\int_0^1 dt\int_0^1 dt'\,f(t,t'),
\end{eqnarray}
\noindent it can be shown by taking the limit $n\to \infty$ that,
\begin{eqnarray}\label{gen}
\Psi( \alpha^{\prime})&=& U(\alpha(1)) \Big[ \mathcal{P}_t 
\left (\exp {\int_0 ^ {1} dt \, 
\mathcal D_ {t}} \right)
\Psi(\alpha)\Big] U(\alpha^{-1}(0)),
\end{eqnarray}
\noindent where,
\begin{eqnarray}
&& \mathcal{P}_t \left (\exp {\int_0 ^ {1} dt \, 
\mathcal D_ {t}} \right)\Psi(\alpha)=\Psi(\alpha)+\int_{0}^{1} 
dt\,  \mathcal D_{t}\Psi(\alpha)+  \int_{0}^{1} dt_1 \int_{0}^{t_1}dt_2
\mathcal D_{t_1}
\mathcal D_{t_2}\Psi(\alpha)+ \nonumber \\ 
&& +\int_{0}^{1}dt_1 \int_{0}^{t_1}dt_2 \int_{0}^{t_2}dt_3 
\mathcal D_{t_1}
\mathcal D_{t_2}\mathcal D_{t_3} \Psi(\alpha)+\dots,
\end{eqnarray}
\noindent and where the notation is
\begin{eqnarray}
U(\alpha(\sigma))&=&\mathcal{P}_t\left(\, \exp \int_0^ 1
 -A_{\mu}(t)\, N^{\mu}(\sigma,t)dt \, \right),
\\ \mathcal D_{t_n}&=& \int_0^1 d\sigma  \,   N^{\mu}
(\sigma,t_n)\, \mathcal D_{\mu} (\sigma,t_n).
\end{eqnarray}
Thus comparing (\ref{gen}) with (\ref{loop}) we identify
\begin{eqnarray}
 U(l) =\mathcal{P}_t \left (\exp {\int_0 ^ {1} dt \, 
\mathcal D_ {t}} \right),
\end{eqnarray}
\noindent which tell us that each deformation may be 
associated to the loop described in the previous section. Furthermore, 
it can be shown that when one considers only spatial curves 
and their loop deformations the operator $D_L$ 
is the loop derivative $\Delta$ introduced in 
\cite{Gambini-loop formulation,Gambini-Pullin}, since both
are curves generators. 

To derive the Bianchi identity let us perform 
six consecutive loop deformations on an arbitrary curve $\gamma$. 
We consider the loop deformations along 
the edges $\delta x_1, \delta x_2, \delta x_3$ of a parallelepiped. 
The functional $\Psi$ of the curve $\gamma$ is affected by 
\begin{eqnarray}
\Psi( \gamma)&=& \Psi \Big(\prod_{n=1}^6 L_n \, o\,\gamma \Big)=
\prod_{n=1}^6 U(L_n) \Psi(\gamma).
\end{eqnarray}
From (\ref{loop deformation}) we may write 
\begin{eqnarray}\label{bianchi}
\Psi( \gamma)&=&  \prod_{n=1}^6 U(\mathcal {D}_{L_n})
 \Psi( \gamma),
\end{eqnarray}  
\noindent  with
\begin{eqnarray}
L_1= \delta x_1 \delta x_2   \delta x^{-1}_1 \delta x^{-1}_2   ,
 \\  L_2= \delta x_3 \delta x_1   \delta x^{-1}_3 \delta x^{-1}_1  ,  
 \nonumber 
    \\L_3=\delta x_2 \delta x_3   \delta x^{-1}_2 \delta x^{-1}_3  ,
     \nonumber\\L_4=\delta x_1\,o\, (\delta x_3 \delta x_2   \delta x^{-1}_3 
     \delta x^{-1}_2) \,o\,
      \delta x^{-1}_1, \nonumber\\L_5=\delta x_2\,o\, (\delta x_1 \delta x_3 
       \delta x^{-1}_1 \delta x^{-1}_3) \,o\,
      \delta x^{-1}_2 , \nonumber\\L_6= \delta x_3\,o\,( \delta x_2 \delta x_1  
       \delta x^{-1}_2 \delta x^{-1}_1) \,o\,
      \delta x^{-1}_3 .\nonumber     
\end{eqnarray}
But first, an intermediate step is required. We need the calculation
of the loop deformation $\mathcal D_{L}$ along the edges $\delta x_1$,
 $\delta x_2$ of a rectangle with one 
vertex in contact with the curve $\gamma$.
Thus, expanding the four point deformations that produces the single 
loop deformation, to second order in the segments we have
\begin{eqnarray}\label{Ricci}
U(\mathcal D_L) \Psi(\gamma)=(\,1+ [\,\mathcal {D}_{\delta x_1},
\mathcal {D}_{\delta x_2}\,]\,)\Psi(\gamma).
\end{eqnarray}   
And by expanding the left side of eq (\ref{Ricci}) we obtain
 $\mathcal D_L=[\mathcal
 {D}_{\delta x_1},\mathcal {D}_{\delta x_2}]$ which is the
  operator version of the Ricci identity.
We also need the same loop deformation when one vertex of 
the rectangle is connected to the curve    
by a retraced segment $\delta x$. It is easy to show that 
\begin{eqnarray}
U(\mathcal D_{\delta x^{-1}\,L\,o\,\delta x}) \Psi(\gamma)= 
(\,1+  U^{-1}_{\delta x} \,[\,\mathcal 
{D}_{\delta x_1},\mathcal {D}_{\delta x_2}\,]\,U_{\delta x} \,)\Psi(\gamma).
\end{eqnarray}   
Using these expressions, the right side of eq (\ref{bianchi})
 turns to be   
\begin{eqnarray}
&& \prod_{n=1}^6 U(\mathcal {D}_{L_n}) \Psi(\alpha)=(\,1-[\,\mathcal 
{D}_{\delta x_1},\mathcal {D}_{\delta x_2}\,]\,)\times
 (\,1-[\mathcal {D}_{\delta x_1},\mathcal {D}_{\delta x_3}\,]\,)\times
 (\,1-[\,D_{\delta x_2},D_{\delta x_3}\,]\, )\nonumber \\
 && \times(\,1+ U^{-1}_{\delta x_3}\, [\,\mathcal {D}_{\delta x_1},
 \mathcal {D}_{\delta x_2}\,] \,
 U_{\delta x_3})   \times(\,1+ U^{-1}_{\delta x_2} \,[\,\mathcal 
 {D}_{\delta x_1},\mathcal {D}_{\delta x_3}\,] \,U_{\delta x_2}\,)
  \nonumber \\ &&\times(\,1+ U^{-1}_{\delta x_1} \,[\,\mathcal 
  {D}_{\delta x_2},\mathcal {D}_{\delta x_3}\,]\, U_{\delta x_1}\,)\,  \Psi(\alpha),
\end{eqnarray}
\noindent and therefore we obtain    
\begin{eqnarray}
\Psi( \alpha)= \Big(\,1+ \,\mathcal {D}_{\delta x_1} [\,\mathcal 
{D}_{\delta x_2},\mathcal {D}_{\delta x_3}\,]+ 
 \,\mathcal {D}_{\delta x_3} [\,\mathcal {D}_{\delta x_1},\mathcal 
 {D}_{\delta x_2}\,]
+ \,\mathcal {D}_{\delta x_2} \,[\mathcal {D}_{\delta x_3},\mathcal 
{D}_{\delta x_1}\,]   \Big)\Psi(\alpha),
\end{eqnarray}
\noindent which gives the Bianchi identity for the point deformation $D_{\mu}(\sigma)$ 
\begin{eqnarray}
 \mathcal {D} _{\alpha}  [\,\mathcal {D}_{\mu},\mathcal {D}_{\nu}\,]+
  \mathcal {D}_{\mu}[\,\mathcal {D}_{\nu},\mathcal {D}_{\alpha}\,]
+ \mathcal {D}_{\nu}[\,\mathcal {D}_{\alpha},\mathcal {D}_{\mu}\,]=0,
\end{eqnarray}
\noindent recall the expression (\ref{func-der}).
\section{Applications of the Path Dependent Formalism}%..........................................Applications
In this section, the path dependent formalism is used to deduce  
the covariant Taylor expansions 
for non Abelian fields. These expansions were developed as part of 
a method of calculation to found the effective action in quantum field theories 
\cite{Barvinsky}. Generalizations that consider supersymmetric expansions
are given in \cite{Kuzenko}. In the derivation 
of these series we follow a geometrical approach 
based on the deformation of open curves instead of using local operators an in
the standard method.
Thus, in order to compare both methods we review in the first subsection 
the standard method and in the second
the path dependent formalism. This geometrical framework could be a 
starting point to explore expansions on curved base manifolds and where 
singularities are present.
\subsection{Standard Derivation}%.....................................................................Standard Method
We follow the original method 
\cite{Barvinsky}, which we extend here to accommodate besides gravitational covariant derivatives
also gauge covariant ones. 

Let us define
the covariant derivative along the path $x(t)$ given by the operator
\begin{eqnarray}
\frac{D}{dt}=\dot x^{\mu}(t)D_{\mu},
\end{eqnarray}
\noindent where $D_{\mu}$ is the usual covariant derivative; 
for a gauge theory we just have to lift the curve 
$x(t)$ to the bundle. Let us write the parallel transport 
equation for the geodesic curve
 $x(t)$ that connects the points $x_1$ and $x_2$,
\begin{eqnarray}\label{geodsic}
\frac{D\dot  x^{\nu}}{dt}=\dot x^{\mu}D_{\mu}\dot x^{\nu}(t)=0.
\end{eqnarray}
Using (\ref{geodsic}), for the scalar function $f(x(t))$ one has for all $n$,
\begin{eqnarray}
\frac{d^nf(x(t))}{dt^n}=\left[D_{\nu_n} 
\dots D_{\nu_1}f(x)\right]_{x=x(t)}\, \dot x^{\nu_1}\dots \dot x^{\nu_n}.
\end{eqnarray}
\noindent Then, considering the expansion of $f(x(t))$ 
\begin{eqnarray}\label{tay}
f(x(t_2))=\sum_{n=0}^\infty \frac{1}{n!}\left[\frac{d^nf(x(t))}
{dt^n}\right]_{t=t_1}(t_2-t_1)^n,
\end{eqnarray}
\noindent and defining, 
\begin{eqnarray}
\sigma^{\mu}(x_1,x_2)=(t_2-t_1)\left[\frac{dx^{\mu}(t)}{dt}\right]_{t=t_1},
\end{eqnarray}
\noindent we arrive to the expression, 
\begin{eqnarray}\label{serie}
f(x_2)=\sum_{n=0}^\infty \frac{1}{n!} \, \sigma^{\nu_1}(x_1,x_2) \dots  \sigma^
{\nu_n}(x_1,x_2)  D_{\nu_n} \dots D_{\nu_1}f(x_1).
\end{eqnarray}

In order to obtain the covariant Taylor series we consider the  
field composed with the two parallel propagators as
$U(x',x)\,
\varphi(x)\, U(x,x') $. Since the composition behaves as a scalar with
 respect to the point $x$ 
we can apply the expansion
(\ref{serie}),
\begin{eqnarray}
U(x'',x')\, \varphi(x')\, U(x',x'')&=&\sum_{n=0}^\infty  \frac{1}{n!} \, 
\sigma^{\nu_1}(x,x') \dots 
 \sigma^{\nu_n}(x,x') \, D_{\nu_n}^x \dots D_{\nu_1}^x\nonumber \\
 && \times ~~U(x'',x)\, \varphi(x)\, U(x,x'').
\end{eqnarray}
\noindent  Now, we need the identity 
\begin{eqnarray}
 \sigma^{\nu_1}(x,x') \dots  \sigma^{\nu_n}(x,x') \, D^x_{\nu_n} 
 \dots D^x_{\nu_1}\, U(x',x)=0,
\end{eqnarray}
\noindent that results by using $\sigma^{\mu}D_{\mu}\sigma^{\nu}= \sigma^{\nu}$ and 
making the operator 
$\sigma^{\mu}D^x_{\mu}$ act successively on
\begin{eqnarray}
\sigma^{\mu}D^x_{\mu}\, U(x',x)=0,
\end{eqnarray}
\noindent We finally obtain the covariant Taylor series for the field $\varphi(x)$,
taking $x''=x'$ and multiplying by $U^{-1}(x,x')$ and $U^{-1}(x',x)$,

\begin{eqnarray}\label{standard}
U(x,x')\, \varphi(x')\, U(x',x)=\sum_{n=0}^\infty  \frac{1}{n!} \sigma^{\nu_1}(x,x') \dots 
 \sigma^{\nu_n}(x,x') \, D_{\nu_n} \dots D_{\nu_1}\, \varphi(x).
\end{eqnarray}
\subsection{Geometrical Derivation}%.......................................................Geometrical Derivation
The deduction in the geometrical approach is rather simple. This consist 
in performing a point deformation on a path dependent field while 
taking the limit where the curve collapse to a point.
To illustrate the idea let us consider them in two separately
deformations, both acting on the path-dependent field 
$\Psi(\,\gamma\,)$ defined to be the
composition of the path propagator $U(\,\gamma\,)$ and the insertion of the 
field $\phi(x)$ as,
\begin{eqnarray}
\Psi(\,\gamma)= U(\gamma_2) \,  \phi
(x) \,
U(\gamma_1),
\end{eqnarray}
\noindent where $\gamma_1$ and $\gamma_2$ are the curves left in
 both sides of the insertion, see Fig 2.
For the first deformation, drawn in dotted lines, we leave the end points $x_1$ y $x_2$
fixed, so we write
\begin{eqnarray}\label{psi}
\Psi( \gamma^{\prime})&=&U(D)\Psi(\gamma).
\end{eqnarray}
\noindent In the next deformation we carry the end points 
over the paths of $\gamma_1$ and $\gamma_2$ to 
coincide on the point 
$x$. This last deformation produces the curve $\gamma''$ made of the two 
straight line 
segments
connecting $x$ to $x'$ and back.
\begin{figure}
\begin{center}
\includegraphics[width=6cm]{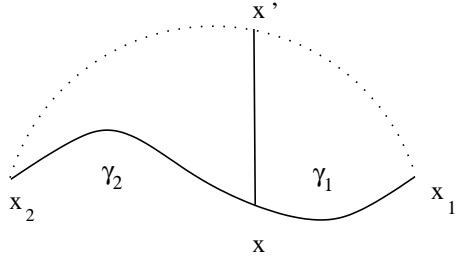}
\caption{Two deformations of the curve $\gamma$.}
\end{center}
\end{figure}
This way the next deformation can be represented (\ref{gen}) by
\begin{eqnarray}
\Psi( \gamma'')&=&U(\gamma^{-1}_2)\left[\, U(D')\Psi(\gamma')\, 
\right]U(\gamma_1^{-1}),
\end{eqnarray}
\noindent and from (\ref{psi}) we have,
\begin{eqnarray}
\Psi( \gamma'')&=& U(\gamma_2^{-1}) \left[\, U(D')\,U(D)\Psi(\gamma)\,
 \right] U(\gamma_1^{-1}).
\end{eqnarray}
Since the effect of the two deformations can be regarded 
as a point deformation 
$D_{\delta x}$, we can write
\begin{eqnarray}
\Psi( \gamma'')&=&U(\gamma_2^{-1})  [\, U(D_{\delta x})  \Psi(\gamma)
\, ]\, U(\gamma_1^{-1}).
\end{eqnarray}
Therefore using the expression
(\ref{Mandelstam}), we obtain covariant Taylor expansions for a
 parallel transported field along a straight line segment $\delta x=x'-x$
\begin{eqnarray}
 U(x,x')\, \phi(x') \,  U(x',x)&=&\sum_{n=0}^\infty  \frac{1}{n!} 
 \delta x^{\nu_1} \dots 
 \delta x^{\nu_n} \, D_{\nu_n} \dots D_{\nu_1}   \phi(x),
\end{eqnarray}
\noindent which coincides with (\ref{standard}), identifying 
$\sigma^{\nu}(x,x')$ with $\delta x^{\nu}$.
\section{Conclusions}%.......................................................................Conclusions
We have defined a path dependent operator in gauge theory, 
which is covariant by construction, and acts by
continuous deformations on the space of smooth 
curves $\Gamma(M)$. The path operator may be seen as generating 
a vector field associated to the deformation of a given curve.
We have adapted the deformation 
to a one-parameter family of diffeomorphism which allows us to define 
the path operator manifestly independent of coordinates.
Although different in nature, since these diffeomorphisms drag many
 points at the same time, we found a close analogy with the Lie 
derivative introduced in general relativity. Therefore, it is in order  
to check whether the path operator $\mathcal D(N)$ belongs to the 
algebra of diffeomorphism or not. 
When we restrict to a sector of $\Gamma(M)$ containing only
spatial curves, their loop deformations (end points fixed) are simply 
the loop 
derivatives defined in \cite{Gambini-loop formulation,Gambini-Pullin}.
We have also established a clear relation between the path derivative 
introduced here and the area 
and end point derivative. This has been done by comparing the action
 of the path derivative with well-known equations 
involving phase factors and path dependent scalar fields. 
We have calculated the finite variation of a functional when its
 argument is changed by successive infinitesimal deformations. This change 
has been interpreted through the action of the group of loops on 
arbitrary paths, which we
 have represented by the action of the covariant path
operators. Geometrically, the path operator is identified with an 
infinitesimal generator of curves.
Ricci and Bianchi
identities have been obtained for loop deformations along the 
edges of a rectangle. 
We have deduced covariant Taylor expansions 
for non Abelian fields by considering the deformation of open curves.
The reason why closed curves cannot be used in the derivation is because
there is no way to deform closed curves to points without enclosing area, which would have
produced loop deformations
instead of point deformations and local covariant derivatives. 
Another feature of the derivation is the fact that we have used a deformation 
that moves all the points 
of the original curve not considered by other path operators. 

It is important to note the close analogies found with the
 approach \cite{Gambini-loop formulation}, not only the identification 
of the loop derivative as a particular case of deformation, but also  
the role of the group of loops as the fundamental geometrical structure 
underlying the gauge theory. 
An immediate main difference with this approach, besides the 
space dimension, is the global action of the 
path derivative. The path derivative action, allows us 
to deform the entire curve 
which is not possible using the loop derivative, since
 its more abstract action
consist in attaching infinitesimal loops on curves. 
\section*{Acknowledgements}
This work was partially supported by CONACYT-40745-F,47211-F. 
The author would like to acknowledge 
useful comments and suggestions from Luis F. Urrutia, Alejandro 
Corichi, and J. David Vergara. I am indebted to 
Hugo A. Morales-Tecotl 
for inviting me at UAMI and for discussions and suggestions on 
possible future applications. A special thanks 
to Stella Huerfano for stimulating conversations on innumerable
 mathematical aspects of this work.
\bibliographystyle{diss}
\bibliography{literature_gravity}

\end{document}